\documentclass[a4paper,11pt]{article}
\pdfoutput=1 

\usepackage{jheppub} 

\usepackage[T1]{fontenc} 

\usepackage{braket}
\usepackage{amsthm}
\usepackage{caption}
\usepackage{subcaption}
\usepackage{listings}

\hypersetup{pdftitle={The Complexity of Finding Ryu-Takayanagi Surfaces}, pdfauthor={Ning Bao, Aidan Chatwin-Davies}, citecolor=blue,linkcolor=blue,urlcolor=blue,citecolor=blue}

\newcommand{\Sec}[1]{section~\ref{#1}}

\newcommand{\Fig}[1]{figure~\ref{#1}}
\newcommand{\App}[1]{appendix~\ref{#1}}

\newtheorem{theorem}{Theorem}
\newtheorem{proposition}[theorem]{Proposition}
\newtheorem{definition}[theorem]{Definition}
\newtheorem{problem}[theorem]{Problem}

\newcommand{\pf}{\noindent {\bf Proof: }}
\newcommand{\eop}{{\hspace*{\fill}$\square$}}

\definecolor{purple}{rgb}{0.5,0,0.5}

\preprint{CALT-TH-2016-023}

\title{\boldmath The Complexity of Identifying Ryu-Takayanagi Surfaces in $\mathrm{AdS}_3/\mathrm{CFT}_2$}

\author{N. Bao}
\author{and A. Chatwin-Davies}
\affiliation{Walter Burke Institute for Theoretical Physics,\\ California Institute of Technology,
Pasadena, CA 91125}

\emailAdd{ningbao@theory.caltech.edu}
\emailAdd{achatwin@caltech.edu}

\abstract{We present a constructive algorithm for the determination of Ryu-Takayanagi surfaces in $\mathrm{AdS}_3/\mathrm{CFT}_2$ which exploits previously noted connections between holographic entanglement entropy and max-flow/min-cut. We then characterize its complexity as a polynomial time algorithm.}

\begin{document} 
\maketitle
\flushbottom

\section{Introduction}
\label{sec:intro}

The calculation of entanglement entropy $S$ is a key aspect in understanding the degree of quantumness of a system.
While this is a problem that is generically difficult for arbitrary quantum systems, Ryu and Takayanagi \cite{Ryu:2006bv} beautifully simplified the calculation for field theories that possess classical gravitational duals \cite{Maldacena:1997re} through their eponymous formula,
\begin{equation}
S(A) = \frac{\mathrm{area}(\tilde A)}{4G_N} \, .
\end{equation}
In the above, $A$ is a region in the boundary conformal field theory, and $\tilde A$ is a minimal-area surface in the bulk gravitational dual such that $\partial A = \partial \tilde A$.
The Ryu-Takayanagi formula essentially translates the abstract algebra question of taking partial traces of density matrices into a geometric one.

A natural question to ask is exactly how difficult is it to use the Ryu-Takayanagi formula to calculate the entanglement entropy of a boundary region?
In arbitrary dimensions, even in the case where $A$ consists of a single simply-connected region, the problem of finding the bulk minimal surface is famously difficult.
It is known, for example, that even a discretized version of the problem is {\bf NP}-hard for a bulk that has three spatial dimensions \cite{Agol:2006}. 
The problem simplifies considerably in $\mathrm{AdS}_3/\mathrm{CFT}_2$, where a spacelike slice through the spacetime results in a one-dimensional boundary and a two-dimensional bulk.
Simply-connected boundary regions are just intervals that are entirely characterized by their two endpoints, and when the bulk is itself simply-connected, the corresponding bulk minimal-area surface is a single geodesic that is anchored on the boundary at the interval's endpoints.
Nevertheless, the story becomes more complicated when $A$ consists of a set of disjoint subregions in the one-dimensional boundary conformal field theory.
The question boils down to finding the correct union of geodesics (which run between boundary subregion endpoints) that is altogether both minimal and homologous to $A$.
For $A$ which is the union of $n$ subregions, the brute force solution consists of checking every combination of $n$ geodesics that run between the $2n$ endpoints---a task that scales exponentially in $n$.
Is there a more efficient way to identify the correct union of geodesics, or is the combinatorics of boundary subregions a source of hardness even in a one-dimensional boundary?

Strikingly, we find that there is a strong simplification to a polynomial time algorithm in three-dimensional gravity.
Following previous inspiring and precise statements of a connection between the Ryu-Takayanagi conjecture and max-flow/min-cut \cite{Freedman:2016zud,Bao:2015bfa}, we first devise a constructive algorithm that reduces the problem of determining the minimal-area surface in the context of $\mathrm{AdS}_3/\mathrm{CFT}_2$ to solving max-flow/min-cut on a graph.
Crucially, the latter problem can be solved in polynomial time.
We then analyze the computational overhead that is required to reduce the problem to max-flow/min-cut and verify that it requires no more than polynomial time as well.

The organization of the paper is as follows: In \Sec{sec:alg} we present the algorithm to identify the bulk minimal surface in mathematical terms, and in \Sec{sec:complex} we analyze the complexity of this algorithm.
In \Sec{sec:topo}, we discuss how our approach generalizes to nontrivial bulk topologies.
Finally, in \Sec{sec:conc} we conclude with a few remarks.

\section{An algorithm to identify minimal-length bulk surfaces}
\label{sec:alg}

We begin by precisely stating the problem.
Let $X$ be a spatial slice of an asymptotically $\mathrm{AdS}_3$ spacetime, and suppose that $X$ is simply-connected.
Further, suppose that $X$ is holographically dual to a CFT state $\rho$ defined on its boundary, $\partial X$.
Let $\{A_i\}_{i=1}^n$ with $n \geq 2$ be a collection of non-empty, simply-connected, closed, disjoint boundary regions, i.e., $A_i \subset \partial X$, $A_i \neq \emptyset$, $A_i = \mathrm{cl}(A_i)$ for $i \in [n]$,\footnote{We use the notation $[n] \equiv \{1, 2, \dots, n\}$ as in \cite{Bao:2015bfa}.} and $A_i \cap A_j = \emptyset$ for $i \neq j$.
What is the Ryu-Takayanagi surface, i.e., the minimal-length bulk surface $\tilde A$ that is homologous to $A = \bigcup_{i=1}^n A_i$?
Or, if there are several minimal-length surfaces, what is one of them?
Note that we may consider strictly disjoint regions without loss of any generality;
two overlapping or touching regions $A_{i}$ and $A_{i+1}$ may be fused, since the area of any surface that subtends $A_i$ and $A_{i+1}$ separately can only be greater than or equal to the area of a surface that subtends $A_i \cup A_{i+1}$. 

\begin{figure}
\centering
\begin{subfigure}[c]{0.48\textwidth}
	\centering
	\includegraphics[width=\textwidth]{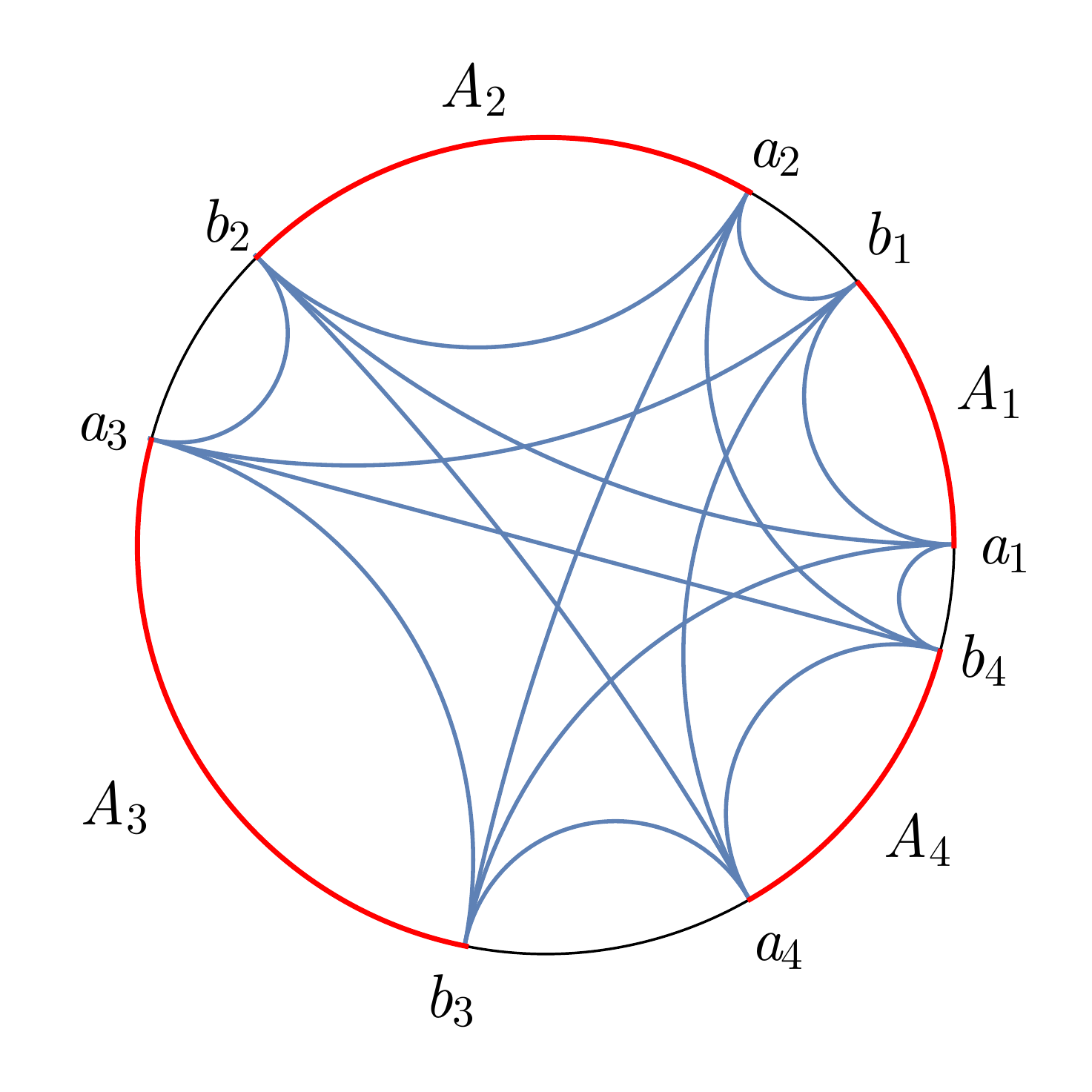}
	\caption{}
	\label{fig:geodesics}
\end{subfigure}
~
\begin{subfigure}[c]{0.48\textwidth}
	\centering
	\includegraphics[width=\textwidth]{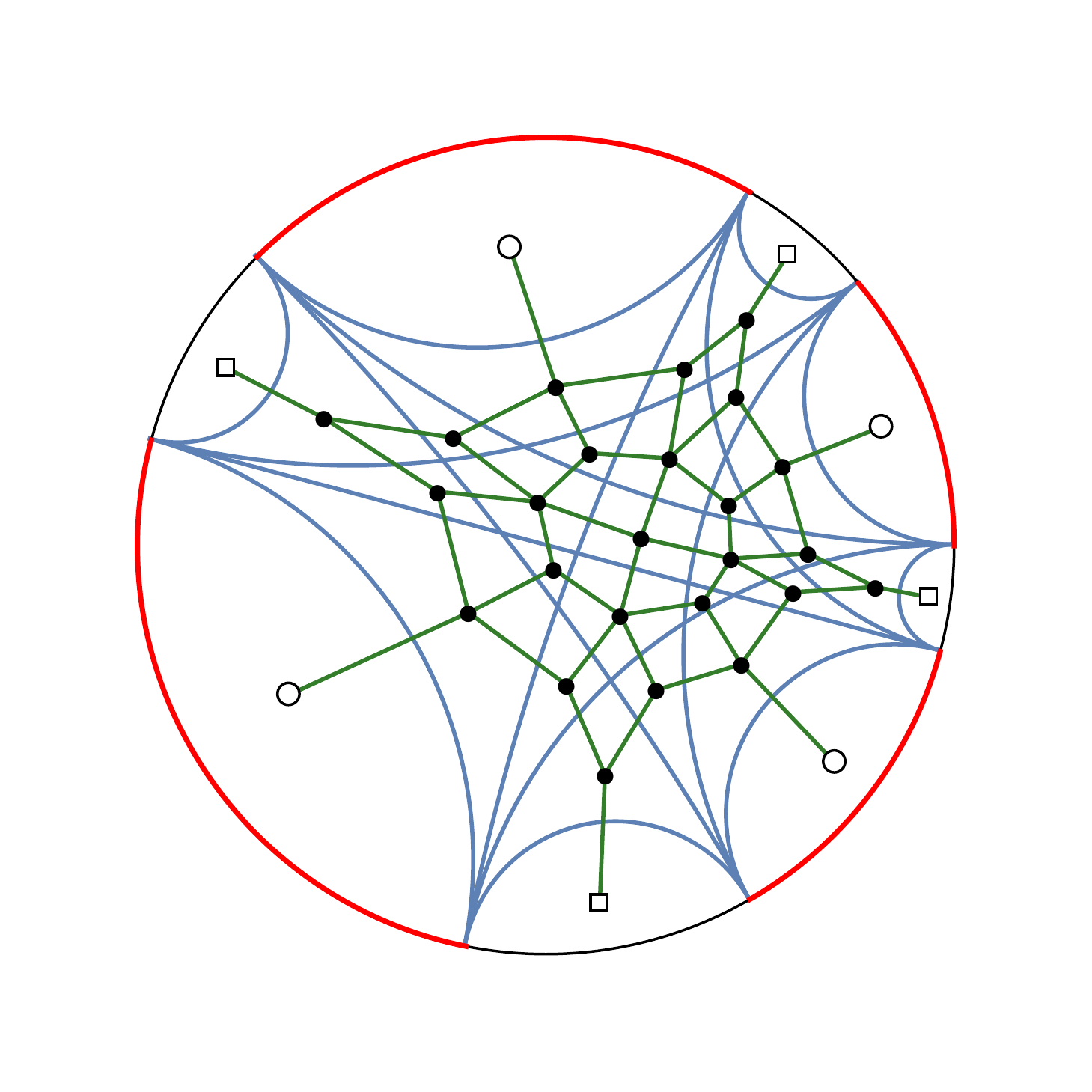}
	\caption{}
	\label{fig:dual}
\end{subfigure}
~
\begin{subfigure}[b]{0.48\textwidth}
	\centering
	\includegraphics[width=\textwidth]{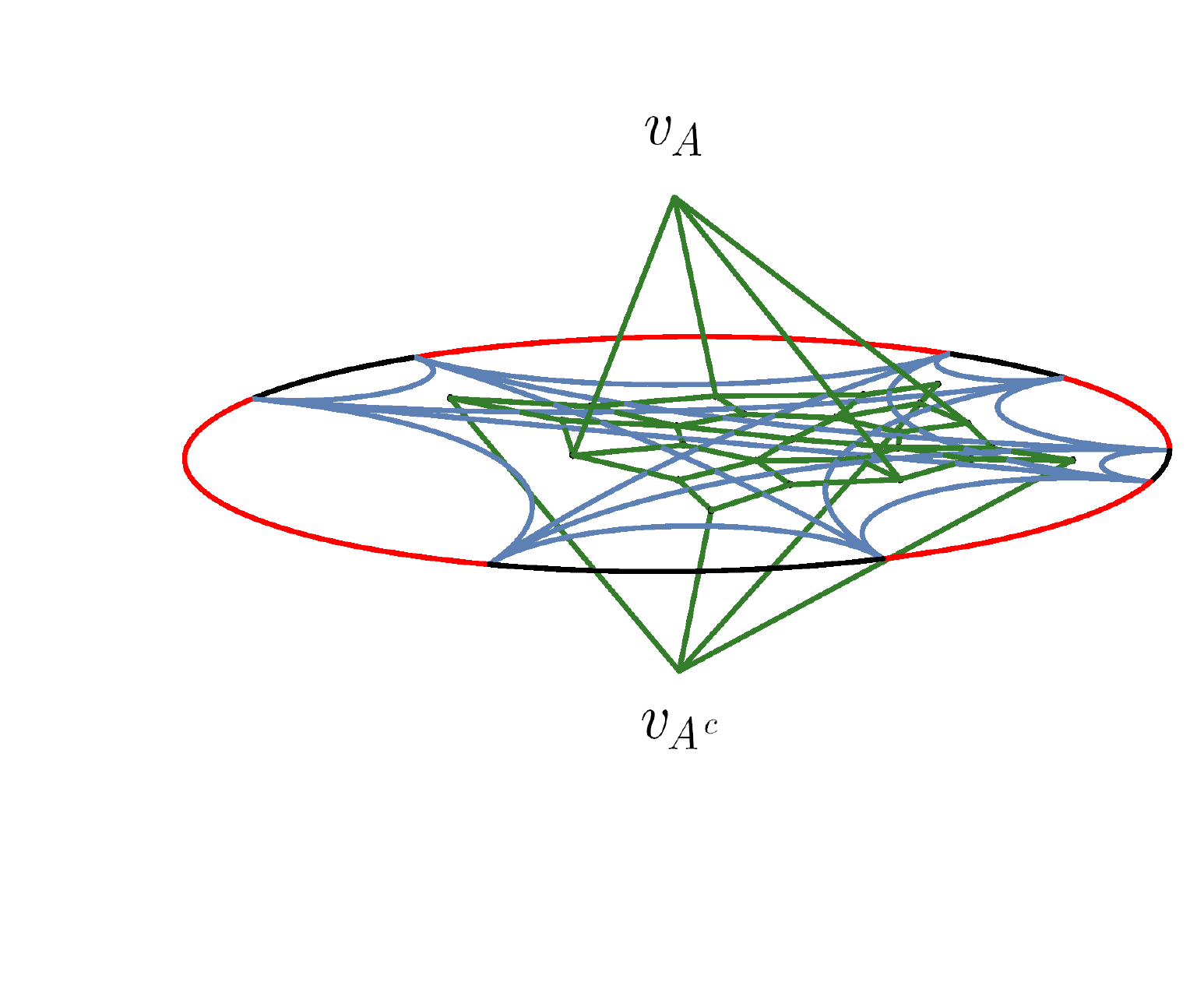}
	\caption{}
	\label{fig:dual3D}
\end{subfigure}
\caption{Variation on the graph construction from \cite{Bao:2015bfa} (cf. their figure 4), illustrated for $n=4$. (a) The boundary regions $A_1, \dots, A_4$ are shown in red and the geodesics which link each $a_i$ and $b_j$ pair, $1 \leq i,j \leq 4$, are shown in blue. (b) The graph $\tilde \Gamma$ is constructed by placing a vertex in each bulk region $X_\alpha$, and vertices are linked when their respective bulk regions share a geodesic segment as an edge. All of the hollow circular nodes are identified as a single vertex $v_A$, and all of the hollow square nodes are identified as a single vertex $v_{A^c}$.
This identification is illustrated in (c).}
\end{figure}

Our approach to identifying $\tilde A$ is to reformulate the question as a problem on a graph using a construction that is a variation on the one presented in section 3 of \cite{Bao:2015bfa}.
Without loss of generality, suppose that the $A_i$ are numbered according to the order in which they appear going counter-clockwise along $\partial X$ and let $A_i = [a_i, b_i]$, again with respect to counter-clockwise ordering.
Next, for each $a_i$, draw geodesics between it and every $b_j$  so that each geodesic subtends a boundary region $[a_i,b_j]$ whose interior contains zero or an even number of boundary endpoints\footnote{The graph formed by the endpoints as vertices and the geodesics as edges is known as a complete bipartite graph.} (\Fig{fig:geodesics}).
Because $X$ is two-dimensional and simply-connected, these geodesics are precisely the curves that could possibly make up the bulk minimal surface, or in other words, $\tilde A$ is a subset of these geodesics.
Since each $a_i$ is connected to each of the $n$ endpoints $b_j$, there are $n^2$ geodesics in total, and since in the minimal surface each endpoint must be connected to only one other endpoint by a geodesic, $\tilde A$ consists of $n$ geodesics.
Therefore, the task of finding $\tilde A$ amounts to identifying $n$ of the $n^2$ geodesics whose cumulative length is minimal, subject to the constraint that they must together subtend $A$.

The set of these geodesics, together with $\partial X$, partition $X$ into a collection of bulk pieces $\{X_\alpha\}$.
Define a weighted graph $\tilde \Gamma$ by placing a vertex $v_\alpha$ in each of these pieces, and connect two vertices $v_\alpha$, $v_{\alpha^\prime}$ with an edge $e_{\alpha\alpha^\prime}$ if the pieces to which they belong share an edge, which is itself a segment of a single geodesic (\Fig{fig:dual}).
Define the weight of $e_{\alpha\alpha^\prime}$ to be the proper length of this geodesic segment, i.e., $\omega(e_{\alpha \alpha^\prime}) = | X_{\alpha} \cap X_{\alpha^\prime} |$, where $|\cdot|$ denotes proper length in this context.
Finally, merge all of the vertices for which $(X_\alpha \cap \partial X) \subset A$ into a single vertex $v_A$, and similarly merge all of the vertices for which $(X_\alpha \cap \partial X) \subset A^c$, where $A^c = \partial X - A$, into another single vertex $v_{A^c}$ (\Fig{fig:dual3D}).

Next, define a cut in the following way\footnote{Note that this definition is slightly different from that of \cite{Bao:2015bfa}.}:

\begin{definition}
A \emph{$k$-cut} $C$ is a subset of the edges of a graph $G$ such that, upon removal of the edges in $C$, $G$ is partitioned into $k$ disjoint connected components.
The weight of the cut, denoted by $|C|$, is defined as the sum of the weights of the edges that constitute the cut, i.e.,
\begin{equation}
|C| = \sum_{e \in C} \omega(e) \, .
\end{equation}
\end{definition}

\noindent We then arrive at the following result:

\begin{proposition}
Let $C^*$ be a minimal-weight 2-cut that separates $v_A$ and $v_{A^c}$ in the graph construction above.
Then, the union of the geodesic segments to which each $e_{\alpha\alpha^\prime} \in C^*$ corresponds is $\tilde A$, i.e.,
\begin{equation}
\tilde A = \bigcup_{e_{\alpha\alpha^\prime} \in C^*} (X_\alpha \cap X_{\alpha^\prime})
\end{equation}
\end{proposition}

\pf
Upon close examination, one can see that the proposition follows from the proof of lemma 3 in \cite{Bao:2015bfa}.
To show this, first recall the definition of the graph from \cite{Bao:2015bfa}.
Their final graph, which we denote by $G$, is constructed out of boundary-anchored geodesics in the same way as $\tilde \Gamma$ (except for the final step where the two sets of boundary vertices are merged); however, the set of geodesics is different.
Namely, only those geodesics which constitute the actual minimal surfaces for all possible unions of subsets of $\{A_i\}_{i=1}^n$ are used.
Explicitly, for all subsets $I \subseteq [n]$, let $A_I = \bigcup_{i \in I} A_i$ and let $\tilde A_I$ be the corresponding Ryu-Takayanagi surface.
Then, $G$ is obtained placing a vertex in each of the pieces into which $X$ is split by $\bigcup_{I \subseteq [n]} \tilde A_I$.
In particular, note that $\bigcup_{I \subseteq [n]} \tilde A_I$ is contained in the set of geodesics that we use to define our graph $\tilde \Gamma$.

Next, recall the content of lemma 3 of \cite{Bao:2015bfa}.
For each $I \subseteq [n]$, define the \emph{discrete entropy} as
\begin{equation}
S^*(I) = \min \frac{|C_I|}{4 G_N} \, ,
\end{equation}
where $G_N$ is Newton's constant and where the minimization is over all $k$-cuts $C_I$ that separate the $|I|$ boundary vertices corresponding to the pieces $X_\alpha$ for which $(X_\alpha \cap \partial X) \subseteq A_I$ from the rest of the graph.
\cite[Lemma 3]{Bao:2015bfa} then states that $S^*(I)$ coincides exactly with the conventional Ryu-Takayanagi entropy $S(I) = |\tilde A_I|/4G_N$, and so it also follows that $|\tilde A_I| = \min |C_I|$.
Moreover, the proof of this lemma further establishes that the minimal cut, call it $C_I^*$, actually corresponds to the Ryu-Takayanagi surface $\tilde A_I$ (or possibly another equivalent surface with the same length if the minimal surface is not unique).

Now, if the two sets of geodesics that are used to define our graph $\tilde \Gamma$ and the graph $G$ from \cite{Bao:2015bfa} had been the same, then the proof would be complete since we would have that $C^* = C_{[n]}^*$.
However, we must establish that the minimal cut $C^*$ actually corresponds to a Ryu-Takayanagi surface, given that the set of geodesics that generates $\tilde \Gamma$ can be larger than the set that generates $G$.
This result follows from the observation that a minimal cut $C_I^*$ in $G$ correctly identifies a Ryu-Takayanagi surface when $I \neq [n]$.
In this case, one may iteratively think of $G$ as being generated by more geodesics than those contained in $\bigcup_{I^\prime \subseteq I} \tilde A_{I^\prime}$.
This is precisely the case for $\tilde \Gamma$, which is itself generated by at least those geodesics which make up $\bigcup_{I \subseteq [n]} \tilde A_{I}$, and so the minimal cut $C^*$ corresponds to $\tilde A$.

\eop

As such, in our graph construction, the problem of finding the Ryu-Takayanagi surface for the collection of boundary regions $A$ corresponds to finding $C^*$, i.e., solving the max-flow/min-cut problem between the two vertices $v_A$ and $v_{A^c}$ on the graph $\tilde \Gamma$.
Importantly, this is a problem that can be solved efficiently in a time that is polynomial in the number of vertices and edges of $\tilde \Gamma$ (see e.g. \cite[Chapter 5.4]{Chartrand:1993}).
Therefore, in order to show that the whole task of finding the Ryu-Takayanagi surface can be completed in polynomial time, all that is left is to establish that the overhead in setting up the graph construction above takes no more than polynomial time in $n$ and that the number of vertices of $\tilde \Gamma$ is no more than polynomial in $n$.

\section{Complexity analysis}
\label{sec:complex}

We now revisit the algorithm presented above and verify that that each step has an algorithmic complexity that is polynomial in the number of boundary regions, $n$.
First, we note that the problem can be restated as a decision problem:

\begin{problem}
Given as input
\begin{itemize}
\item[i.] a Riemannian metric $g_{ij}(x)$ together with a coordinate ultraviolet cutoff $\Lambda$ that describes a simply-connected, asymptotically-hyperbolic, two-dimensional manifold $X$,
\item[ii.] a list of $n$ pairs of points on the conformal boundary of $X$, $\{[a_i,b_i]\}_{i=1}^n$, that specify $n$ non-empty, disjoint, closed, simply-connected intervals in $\partial X$, and
\item[iii.] a permutation $\sigma: [n] \rightarrow [n]$ that identifies $n$ geodesics that connect $a_i$ with $b_{\sigma(i)}$ for $i \in [n]$ and that together subtend the intervals $[a_i,b_i]$,
\end{itemize}
does there exist another permutation $\sigma^\prime$ such that
\begin{equation}
\sum_{i=1}^n |\gamma_{i,\sigma^\prime(i)}|_\Lambda < \sum_{i=1}^n |\gamma_{i,\sigma(i)}|_\Lambda
\end{equation}
up to a numerical precision $\epsilon$, where $\gamma_{i,\sigma^\prime(i)}$ denotes the geodesic between $a_i$ and $b_{\sigma(i)}$ and $|\gamma_{i,\sigma^\prime(i)}|_\Lambda$ is its proper length with the cutoff $\Lambda$ in place?
(Assume that the $a_i$ and $b_i$ have enough digits of precision to compute at the global precision $\epsilon$.)
\end{problem}

\noindent Of course, the decision problem can be answered by carrying out the algorithm above to actually find the minimal surface.
One reason for writing out this restatement it to clearly identify two sources of algorithmic complexity: complexity in $n$, the number of boundary regions, as well as the numerical complexity that is a consequence of having to compute real-valued geometric quantities up to precision $\epsilon$.
We will focus on the complexity in $n$, but it should be understood that the overall complexity has some multiplicative scaling $O(f(\epsilon))$ which depends on the numerical techniques that one uses to compute geometric quantities.

\subsection*{A dual for a dual}

\begin{figure}[h]
\centering
\includegraphics[scale=0.5]{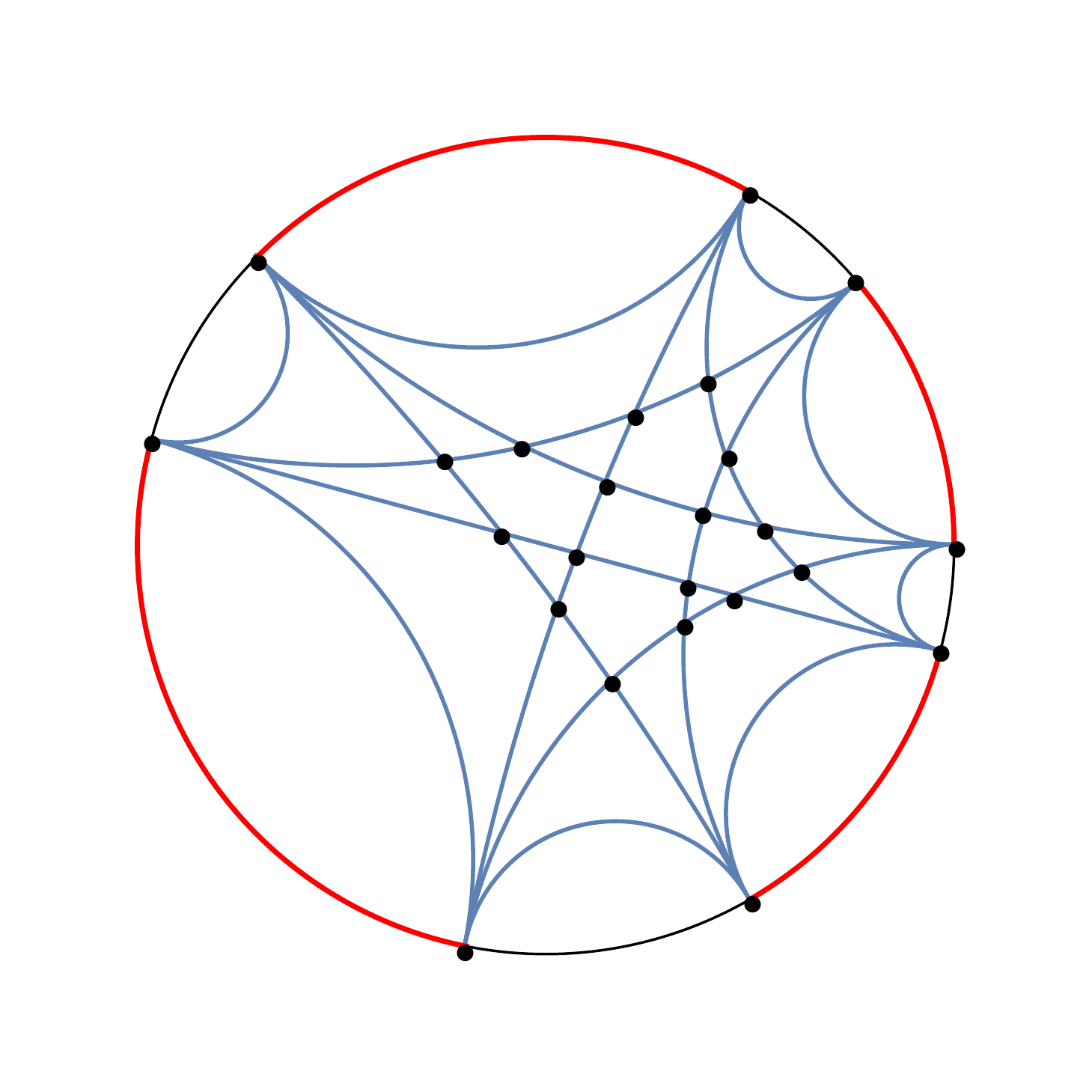}
\caption{The graph $\Gamma$ is constructed by placing a vertex at each point where two or more geodesics intersect and at the boundary interval endpoints $a_i$ and $b_j$. Vertices are connected by the geodesic and boundary segments on which they lie.}
\label{fig:GammaGraph}
\end{figure}

As a preliminary step, it is useful to define a second graph, $\Gamma$, by placing a vertex at every point where two or more geodesics intersect and at each of the $a_i$ and $b_i$.
Connect two vertices with an edge if they are adjacent to each other on a single geodesic, and also add an edge in between each adjacent boundary endpoint (so that the vertices at $a_1$ and $b_1$, at $b_1$ and $a_2$, at $a_2$ and $b_2$, etc. gain an additional edge connecting them).
With this definition, $\tilde \Gamma$ is (up to the merger of the boundary vertices into $v_A$ and $v_{A^c}$) the dual graph of $\Gamma$, which will be useful for counting (\Fig{fig:GammaGraph}).

\subsection*{Finding the geodesics}

Since $X$ is two-dimensional, any boundary-anchored geodesic can be parametrized by two real numbers, for instance, its two endpoints on $\partial X$.
As such, drawing the $n^2$ geodesics between the $a_i$ and the $b_j$ consist of solving the geodesic equation in terms of the two free parameters and then listing the $n^2$ specific solutions.
When the geodesic equation has a closed-form solution, obtaining each geodesic is an $O(1)$ overhead.
For example, when $X$ is the hyperbolic plane, it is straightforward to show that geodesics in the Poincar\'e disk are circular arcs that are normal to the boundary, and a specific arc can be labelled by its endpoints.
Here, one must only solve the geodesic equation once with its endpoints as free parameters.
If the geodesic equation does not have a closed-form solution and/or one works numerically, then constructing and digitally representing each geodesic will have some complexity that depends on $\epsilon$.
Note, however, that this is independent of the number of boundary regions, $n$.
Therefore, the scaling of this step is $O(n^2)$.
We will suppose that the output of this subroutine is a list of functions $\gamma_{i,j} : [0,1] \rightarrow X$ that parametrize the geodesics, i.e., $\gamma_{i,j}(0) = a_i$, $\gamma_{i,j}(1) = b_j$, and whose runtime is independent of $n$.

\subsection*{Number of vertices in $\Gamma$}

Let $V$ be the number of vertices in $\Gamma$.
This is equal to $2n$ (the number of boundary region endpoints) plus the number of times that the $n^2$ geodesics intersect each other.
We can upper bound the latter quantity by $p \cdot \binom{n^2}{2}$, which would be the number of intersections if every geodesic intersected every other geodesic at most $p$ times.
We must assume that $p$ is a bounded constant.
Generically, we expect that $p=1$; in the case where $X$ is simply-connected and has nonpositive curvature everywhere, then $p=1$ is implied by the Cartan-Hadamard Theorem, which guarantees that pairs of points are connected by a unique geodesic (see, for instance, \cite[Theorem 4.5]{Ballmann:1995}).
This point is further discussed in \App{app:intersections}.
With this assumption on $p$, we have that
\begin{equation}
V \leq 2n + p \binom{n^2}{2} = 2n + \tfrac{1}{2} p n^2(n^2-1) = O(n^4) \, .
\end{equation}

\subsection*{Number of edges in $\Gamma$ and $\tilde \Gamma$}

Let $E$ be the number of edges in $\Gamma$.
This is equal to $2n$ (the number of edges that lie on $\partial X$) plus the number of geodesic segments in the interior of $X$.
Note that since $\tilde \Gamma$ is, up to the vertex mergers, the dual graph of $\tilde \Gamma$, this latter quantity is also the number of edges in $\tilde \Gamma$, which we denote by $\tilde E$.

Similarly to the counting of vertices above, the largest number of interior edges is upper bounded by the number of interior edges in a configuration where each geodesic is intersected $p$ times by the $n^2-1$ other geodesics in distinct locations.
In this case, each geodesic is divided into $p(n^2-1) + 1$ segments, and so
\begin{equation}
\tilde E \leq n^2\left[p(n^2-1) + 1\right] = O(n^4) \quad \mathrm{and} \quad E \leq 2n + n^2\left[p(n^2-1) + 1\right] = O(n^4) \, .
\end{equation}

\subsection*{Number of vertices in $\tilde \Gamma$}

The number of vertices in $\tilde \Gamma$ is the number of faces in $\Gamma$ (or equivalently the number of pieces $X_\alpha$), which we denote by $F$, less $2(n-1)$ to account for the vertices that are merged into $v_A$ and $v_{A^c}$.
Since $\Gamma$ is a planar graph, we can use its Euler characteristic to bound $F$.
From $V-E+F=2$, it follows that
\begin{equation}
F=2+E-V \leq 2 + E = n^2\left[p(n^2-1) + 1\right] + 2n + 2 = O(n^4) \, .
\end{equation}
As such, the number of vertices and edges in $\tilde \Gamma$ altogether scales like $O(n^4)$.

\subsection*{Connectivity of the vertices and edge weights}

So far we have established that the size of $\Gamma$ and $\tilde \Gamma$ scales like $O(n^4)$, but we must also establish that the graphs can be constructed in a number of steps that is polynomial in $n$.
In other words, we must be able to locate vertices, determine their connectivity, and compute edge weights efficiently.

Roughly, locating vertices in $\Gamma$ amounts to checking if each pair of geodesics intersects, where each check is a constant overhead if closed-form solutions for the geodesics are known, or some $\epsilon$-dependent overhead if one works numerically.
This task scales like $\binom{n^2}{2} = O(n^4)$.
Then, as noted above, the faces of $\Gamma$ are the vertices of $\tilde \Gamma$.
The weight of an edge in $\Gamma$ (and also $\tilde \Gamma$) is given by the proper length of its corresponding geodesic segment, and so computing this weight amounts to performing a line integral along the geodesic segment.
At worst, if the exact antiderivative is unknown, evaluating this integral numerically up to a fixed numerical accuracy is again a computational task that must be performed less than $\tilde E$ times, and so the algorithmic complexity of this step scales like $O(n^4)$.

To be a bit more concrete, let us sketch an algorithm to construct a digital representation of $\Gamma$.
Represent $\Gamma$ with a $V \times V$ upper-triangular matrix $M$, and denote the vertices of $\Gamma$ by $w_\alpha$.
For $\alpha < \beta$, the entries of $M$ will be $M_{\alpha\beta} = -1$ if $w_\alpha$ and $w_\beta$ are adjacent vertices on the boundary, $M_{\alpha\beta} = \omega(e_{\alpha\beta})$ if $w_\alpha$ and $w_\beta$ are connected via a shared geodesic segment, and zero otherwise.
Let the first $2n$ vertices be the boundary vertices, i.e., $w_{2\alpha-1} \equiv a_\alpha$ and $w_{2\alpha} \equiv b_\alpha$ for $1 \leq \alpha \leq n$.
For each geodesic $\gamma_{i,j}$, $1 \leq i,j \leq n$, we will construct a list $L_{i,j}$ whose entries are pairs $(w_\alpha,s_\alpha)$ which identify the vertices $w_\alpha$ that lie on $\gamma_{i,j}$, as well as the $s_\alpha \in [0,1]$ which specifies the location $\gamma_{i,j}(s_\alpha)$ in $X$ (and hence also on the geodesic itself) where the vertex lies.
Each $L_{i,j}$ can therefore be initialized with two elements,
\begin{equation}
L_{i,j} = \left\{(w_i,0), (w_{j+1},1) \right\} \, .
\end{equation}
The following pseudo-code then sketches how to construct the $L_{i,j}$ and $M$.
A bold index will denote a composite index, i.e., $\mathbf{i} \equiv i,j$.

\begin{lstlisting}[
  mathescape,
  columns=fullflexible,
  basicstyle=\fontfamily{lmvtt}\selectfont,
]
% First fill in $M$ for boundary vertices
for $\lambda$ from 1 to 2n-1
   $M_{\lambda \, \lambda+1}$ = -1
end for
$M_{1 \, 2n}$ = -1

% Next fill in the interior vertices
$\kappa = 2n+1$
for $\mathbf{i}$ from 1 to $n^2$
   % Build up the $L_\mathbf{i}$
   for $\mathbf{j} > \mathbf{i}$
      if $\gamma_\mathbf{i}$ and $\gamma_\mathbf{j}$ intersect
         label this vertex $w_\kappa$
         find the intersection location $s_\kappa^{(\mathbf{i})}$ on $\gamma_\mathbf{i}$ and $s_\kappa^{(\mathbf{j})}$ on $\gamma_\mathbf{j}$
         append $(w_\kappa,s_\kappa^{(\mathbf{i})})$ to $L_\mathbf{i}$ and $(w_\kappa,s_\kappa^{(\mathbf{j})})$ to $L_\mathbf{j}$;
         $\kappa$++
      end if
   end for
   sort $L_\mathbf{i}$ according to increasing $s_\alpha$
   
   % Compute edge weights and fill in $M$
   for $k$ from 1 to (Length[$L_\mathbf{i}$]-1)
      let $(w_{\alpha_k},s_{\alpha_k})$ be the $k^\mathrm{th}$ element of $L_\mathbf{i}$
      compute the length of $\gamma_\mathbf{i}$ from $s_{\alpha_k}$ to $s_{\alpha_{k+1}}$, i.e., $\omega(e_{\alpha_k \alpha_{k+1}})$
      $M_{\alpha_k \alpha_{k+1}} = \omega(e_{\alpha_k \alpha_{k+1}})$
   end for
end for
         
\end{lstlisting}

\noindent Note that the nested loop beginning on line 9 executes $O(n^4)$ times as expected.
The steps that may contribute numerical $\epsilon$-dependent overhead occur on lines 14 and line 24.
Finally, the various array accesses and other tasks (such as the sorting operation on line 19) will only contribute a polynomial number of steps.
{\begin{center} $\sim$ \end{center}}
We therefore ultimately find that the time it takes to set up our graph construction scales like $O(n^4)$, and that $\tilde \Gamma$ itself has a number of vertices and a number of edges that are each $O(n^4)$.
The complexity of max-flow/min-cut is $O(\tilde E \tilde V^2)$ \cite{Chartrand:1993}, and so the overall complexity of our algorithm is $O(n^{12})$.
As such, the algorithmic complexity of finding the Ryu-Takayanagi surface for $n$ boundary regions is $\mathrm{poly}(n)$ as claimed.

\section{Other bulk topologies}
\label{sec:topo}

The algorithm as described above applies to simply-connected bulk geometries.
In situations where the bulk is topologically nontrivial, there is a new parameter which the algorithm could scale with, namely, the genus $q$ of the bulk topology.
This is because it is no longer true that the minimal surface that is homologous to a single simply-connected boundary region is necessarily made up of a single geodesic.
Consequently, the number of geodesics changes from $n^2$ to some $O(f(q)n^2)$.
But, once all of the possible minimal surfaces are determined, one can simply continue apace with the max-flow/min-cut algorithm as before.

\begin{figure}
\centering
\begin{subfigure}[c]{0.24\textwidth}
	\centering
	\includegraphics[width=\textwidth]{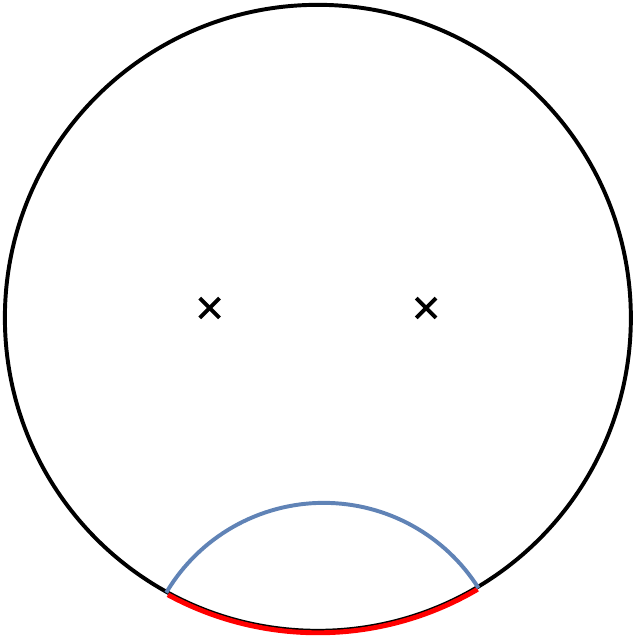}
	\caption{}
	\label{fig:topo1}
\end{subfigure}
~~~~
\begin{subfigure}[c]{0.24\textwidth}
	\centering
	\includegraphics[width=\textwidth]{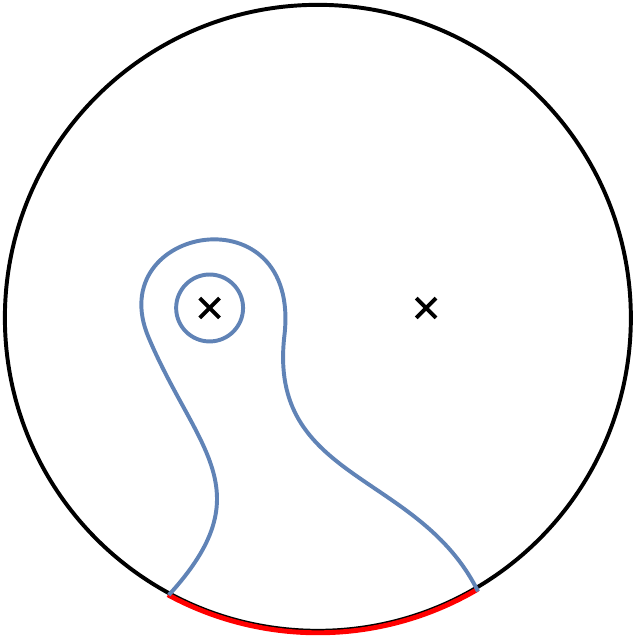}
	\caption{}
	\label{fig:topo2}
\end{subfigure}
~~~~
\begin{subfigure}[c]{0.24\textwidth}
	\centering
	\includegraphics[width=\textwidth]{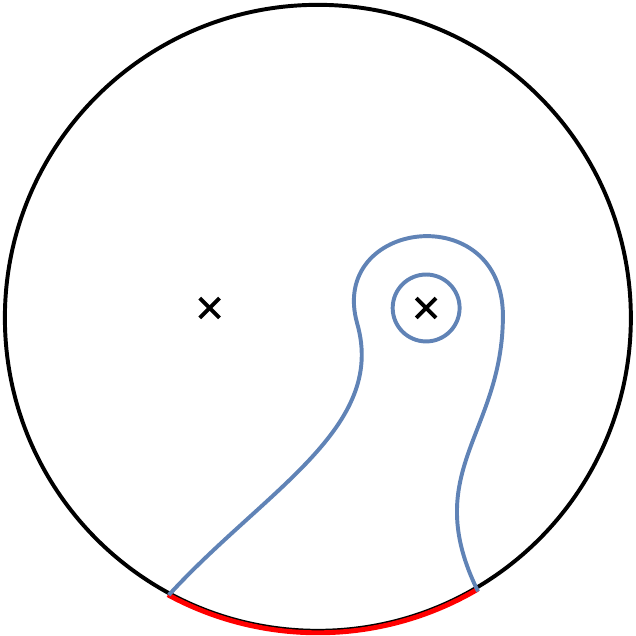}
	\caption{}
	\label{fig:topo3}
\end{subfigure}
~~~~
\begin{subfigure}[c]{0.24\textwidth}
	\centering
	\includegraphics[width=\textwidth]{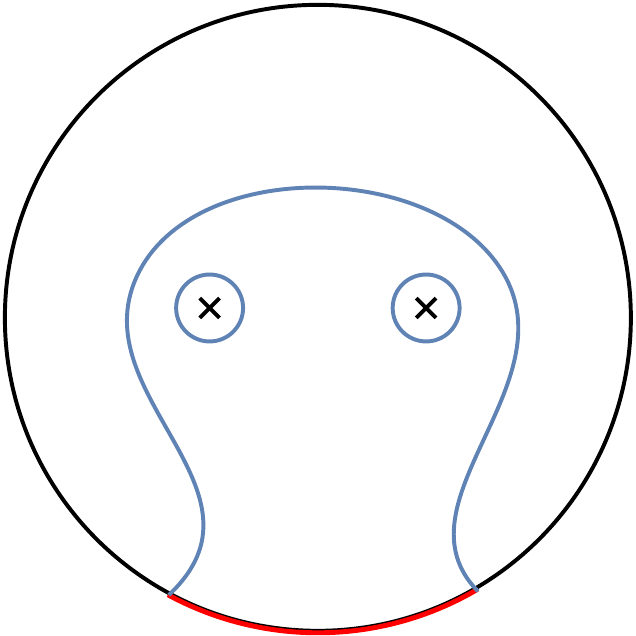}
	\caption{}
	\label{fig:topo4}
\end{subfigure}
~~~~
\begin{subfigure}[c]{0.24\textwidth}
	\centering
	\includegraphics[width=\textwidth]{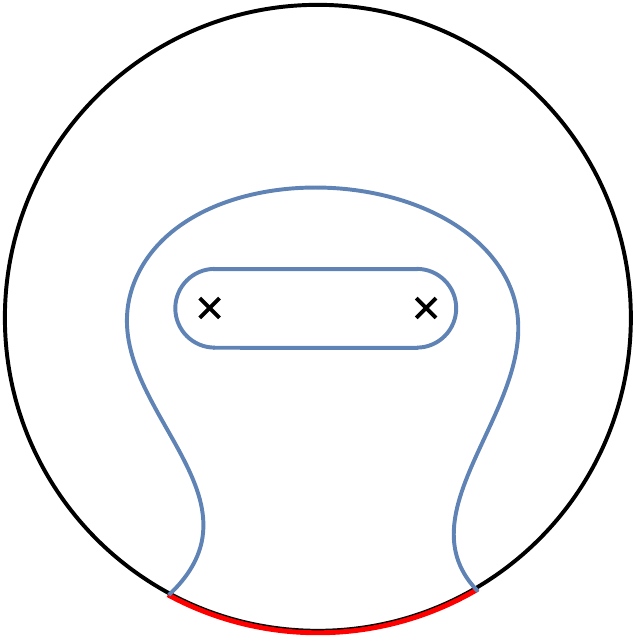}
	\caption{}
	\label{fig:topo5}
\end{subfigure}
\caption{Possible ways that the extremal surface for a single simply-connected boundary region could include or exclude punctures, illustrated for $q=2$. The boundary interval is shown in red, the extremal curve is shown in blue, and puncture are represented by crosses.}
\label{fig:topologies}
\end{figure}

The scaling with $q$ will not change the scaling with the number of boundary intervals; it enters as another independent multiplicative factor.
As a last discussion item, let us estimate what the worst scaling with $q$ could be.
First, consider drawing an extremal path $\mathcal{C}$ between two boundary endpoints $a_i$ and $b_j$ when there are $q$ punctures in $X$ (\Fig{fig:topologies}).
In theory, provided that the geodesics exist, we could choose to include anywhere from zero up to all $q$ of the punctures in $\mathrm{int}([a_i,b_j] \cup \mathcal{C})$, where $\mathrm{int(\cdot)}$ denotes the interior of a closed curve.
Then, noting that there may be up to $\binom{q}{k}$ ways to include $k$ punctures, we identify up to $\sum_{k=0}^q \binom{q}{k} = 2^q$ geodesics in this way.
However, each time that a puncture is included in $\mathrm{int}([a_i,b_j] \cup \mathcal{C})$, we also must draw a geodesic around the puncture so that the total (multiply-connected) extremal curve is homologous to $[a_i,b_j]$.
So, we must also consider the set of all geodesics that enclose anywhere from one to all $q$ of the punctures, where there are $\binom{q}{k^\prime}$ ways to enclose $k^\prime$ punctures.
This gives us another $\sum_{k^\prime=1}^q \binom{q}{k} = 2^q - 1$ geodesics.
Repeating this analysis for every pair of points $a_i$ and $b_j$, we conclude that there are at most $2^q n^2$ geodesics that connect boundary endpoints and $2^q-1$ geodesics that enclose punctures (which remain the same for every pair of boundary endpoints).
Therefore, there will be at most $2^q n^2 + 2^q - 1$ geodesics that seed the rest of the algorithm.
As expected, the scaling in $n$ is unchanged, but the scaling in $q$ can be very large indeed.

\section{Conclusion}
\label{sec:conc}

We have demonstrated that the task of finding the Ryu-Takayanagi surface in three-dimensional gravity for a collection of $n$ boundary subregions has polynomial complexity by exhibiting an algorithm that completes this task in $\mathrm{poly}(n)$ time.
The algorithm consists of converting the geometric problem into a graph-theoretic problem whose solution is given by the max-flow/min-cut between two vertices on a graph.
The procedure is essentially a discretization of Freedman and Headrick's bit thread model \cite{Freedman:2016zud} with the flow being between the boundary subregions $\bigcup_{i=1}^n A_i \equiv A$ and the rest of the boundary, $\partial X - A$.

It should be noted that performing this calculation holographically in many ways complements the computation of entanglement entropy using only $\mathrm{CFT}_2$ techniques, e.g. \cite{Calabrese:2004eu, Holzhey:1994we}.
On one hand, results for simply-connected boundary regions such as the Cardy-Calabrese formula $S=\frac{c}{3} \log \frac{l}{\Lambda}$ do not extend to multiply-disjoint regions.
On the other hand, entanglement entropies of simply-connected regions are the ``primitives'' in our algorithm, and so in pure $\mathrm{AdS}_3/\mathrm{CFT}_2$ for example, one can use the Cardy-Calabrese formula to avoid finding geodesics altogether.
By extension, if you knew how to compute $S$ for simply-connected regions in the boundary for some given holographic $\mathrm{CFT}_2$ state, then assuming the Ryu-Takayanagi conjecture, computing geodesics becomes unnecessary since geodesic length is automatically given by $4GS$.
See, for example, \cite{Maxfield:2014kra}, which demonstrates powerful algebraic methods to compute these geodesic lengths in a broad class of CFT states.


Some future interesting directions would be to use our line of reasoning to clearly delineate the sources of complexity that make the higher dimensional case {\bf NP}-Hard.
For example, it is plausible that the combinatorial aspect of the problem is in general not difficult, but rather that the difficulty arises from the fact that simply-connected boundary regions do not have a canonical shape in higher dimensions.
In a related way, it would also be interesting to extend the algorithm to higher dimensions, but where boundary subregions are restricted to only have certain shapes, e.g., filled $S^{d-1}$ spheres on a $S^d$ conformal boundary.
This sort of setting is important for holographic derivations of the Einstein equations \cite{Lashkari:2013koa,Faulkner:2013ica,Swingle:2014uza} among other applications.

It has also been pointed out that the relationship of holographic entanglement entropy to max-flow/min-cut may extend to covariant formulations \cite{HeadrickHubeny, Freedman:2016zud, Hubeny:2007xt, Wall:2012uf}.
If this is fully established, it would certainly be interesting to see whether our analyses can be extended past minimal surfaces to maximin formulations to arrive at a similar style of conclusion.

\acknowledgments

We thank Adam Bouland, Wilson Brenna, Matthew Headrick, John Preskill, and Michael Walter for helpful discussions and suggestions.
This material is based upon work supported in part by the following funding sources:
N.B. is supported in part by the DuBridge Postdoctoral Fellowship, by the Institute for Quantum Information and Matter, an NSF Physics Frontiers Center (NFS Grant PHY-1125565) with support of the Gordon and Betty Moore Foundation (GBMF-12500028).
A.C.-D. is supported by the NSERC Postgraduate Scholarship program and by the Gordon and Betty Moore Foundation through Grant 776 to the Caltech Moore Center for Theoretical Cosmology and Physics.
This work is supported by the U.S. Department of Energy, Office of Science, Office of High Energy Physics, under Award Number DE-SC0011632.

\appendix
\section{Intersections of geodesics}
\label{app:intersections}

Here we consider the question of how many times two distinct boundary-anchored geodesics in $X$ can intersect.
When $X$ has nonpositive curvature everywhere and is simply-connected, such as the case where $X$ is the hyperbolic plane, then the Cartan-Hadamard theorem implies that every pair of points is connected by a unique geodesic.
We can use this fact to obtain the following result:

\begin{proposition}
Let $X$ by a simply-connected Riemannian manifold with nonpositive curvature everywhere. Then, two distinct geodesics can intersect each other at most once.
\end{proposition}

\pf
We establish the proof by contradiction.
Let $\mathcal{C}_1$ and $\mathcal{C}_2$ be two distinct geodesics, and suppose that they intersect more than once.
Let $p_1$ and $p_2$ be two intersection points, and denote the segment of $\mathcal{C}_1$ (resp. $\mathcal{C}_2$) that connects $p_1$ and $p_2$ by $\mathcal{S}_1$ (resp. $\mathcal{S}_2$).
The lengths of $\mathcal{S}_1$ and $\mathcal{S}_2$ cannot be the same.
This is because the Cartan-Hadamard theorem holds, and so the geodesic that connects $p_1$ and $p_2$ is unique.
Without loss of generality, suppose that $|\mathcal{S}_1| < |\mathcal{S}_2|$.
But then, $(\mathcal{C}_2 - \mathcal{S}_2) \cup \mathcal{S}_1$ is shorter than $\mathcal{C}_2$, which contradicts the assumption that $\mathcal{C}_2$ is a geodesic.

\eop

\noindent Note that this proposition does not exclude the case where $\mathcal{C}_1$ and $\mathcal{C}_2$ overlap on a finite interval.
However, such behaviour does not change the scaling of the number of vertices in $\Gamma$ if we only place vertices at the points where the geodesics first meet.
Also note that the result holds for Riemannian manifolds $X$ of any dimension, and $X$ can be relaxed to a metric space if the curvature is taken to be Alexandrov curvature \cite{Ballmann:1995}.

\begin{figure}[h]
\centering
\includegraphics[scale=0.5]{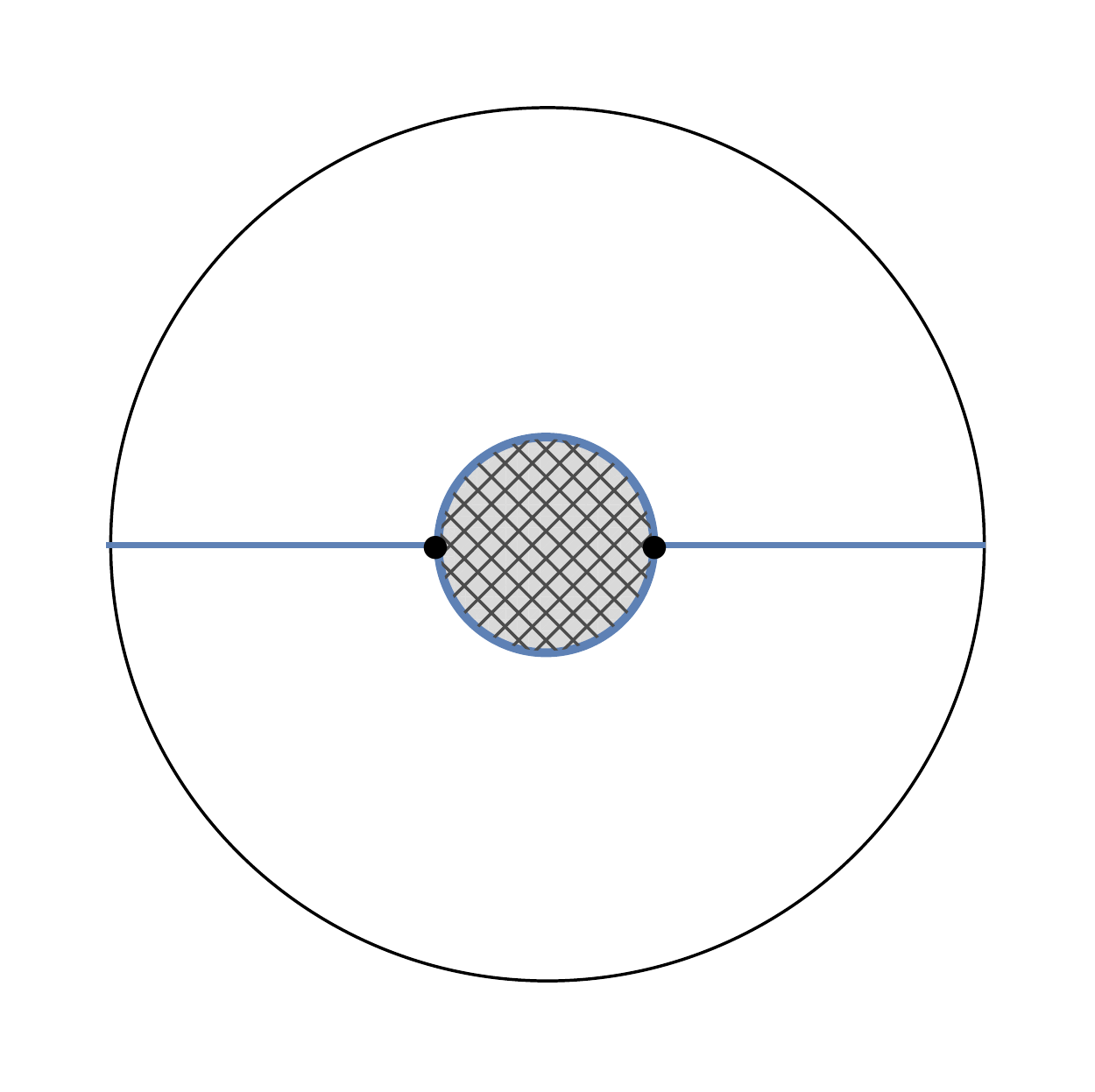}
\caption{Spatial slice through the $(1+2)$-dimensional BTZ spacetime. Two antipodal points (shown in black) on the black hole (the hatched disk) are connected by two distinct geodesics (shown in blue).}
\label{fig:btz}
\end{figure}

Cases where two boundary-anchored geodesics can intersect more than once are necessarily cases where there exist points in $X$ such that the geodesic connecting them is not unique.
An example of such a configuration occurs when $X$ is a slice of the BTZ black hole spacetime \cite{Banados:1992wn} (and is therefore not simply-connected).
Two points in the bulk that are antipodal with respect to the black hole are connected by geodesics of the same length that wrap around either side of the black hole (\Fig{fig:btz}).
The boundary-anchored geodesics on which the two points lie share their boundary endpoints, however, and so this particular configuration is excluded from the configurations that we consider.
As such, it seems reasonable to expect that cases where two boundary-anchored geodesics can intersect more than once and which are allowed by the problem under consideration, if they exist, are highly pathological and could be excluded with an appropriate generic condition.


\end{document}